\documentclass[aps,twocolumn,showpacs]{revtex4}

\usepackage{graphicx}
\usepackage{dcolumn}
\usepackage{bm}

\begin{document} 


\title[]{Evolution to the equilibrium in a dissipative and time dependent 
billiard}

\author{Marcus Vin\'icius Camillo Galia}
\affiliation{ 
Departamento de F\'isica, UNESP - Univ. Estadual Paulista - Av. 24A, 1515 - Bela 
Vista - 13506-900 - Rio Claro - SP - Brazil}%

\author{Diego F. M. Oliveira}
\affiliation{Center for Complex Networks and Systems Research, School of 
Informatics and Computing, Indiana University, Bloomington - Indiana, USA}%

\author{Edson D.\ Leonel}
\affiliation{Departamento de F\'isica, UNESP - Univ. Estadual Paulista - Av. 
24A, 1515 - Bela Vista - 13506-900 - Rio Claro - SP - Brazil}%
\affiliation{Abdus Salam International Center for Theoretical Physics, Strada  
Costiera 11, 34151 Trieste, Italy}

\date{\today}

\begin{abstract}
We study the convergence towards the equilibrium for a dissipative and 
stochastic time-dependent oval billiard. The dynamics of the system is described 
by using a generic four dimensional nonlinear map for the variables: the angular 
position of the particle, the angle formed by the trajectory of the particle 
with the tangent line at the position of the collision, the absolute velocity of 
the particle, and the instant of the hit with the boundary. The dynamics of the 
stationary state as well as the dynamical evolution towards the equilibrium is 
made by using an ensemble of non interacting particles. Finally, we make a 
connection with the thermodynamic by using the energy equipartition theorem.
\end{abstract}

\maketitle

{\bf We investigate the effects of a stochastic perturbation in a dissipative 
time-dependent oval billiard. The system is described in terms of a 
four-dimensional nonlinear mapping for the variables, the angular position of 
the particle, the angle formed by the trajectory of the particle with the 
tangent line at the position of the collision, the absolute velocity of the 
particle, and the instant of the hit with the boundary. Statistical 
investigations for an ensemble of non interacting particles starting in the 
regime of low energy are considered. We show that under the presence of 
dissipation the unlimited energy growth is suppressed. We obtain the average 
velocity analytically and we show its dependence on the dissipation parameter 
confirming a smooth transition from a power law growth to a constant plateau. 
Finally, by using the average squared velocity, we make a connection with the 
thermodynamic and we show that while the temperature grows as a power law and 
reaches a constant plateau, the entropy increases monotonically with the 
energy.}

\section{Introduction}
\label{sec1}


Classical billiards are dynamical system in which a particle, or  an ensemble 
of non-interacting particles, move confined to and  experience collisions with a 
boundary \cite{ref1,ref2,ref3,ref4,add1,ref5,ref6,ref7,ref8}. Basically, they 
are settled in three classes, namely (i) integrable, (ii) ergodic, and (iii) 
mixed. In case (i), the phase space consists of invariant tori filling the 
entire phase space. In case (ii),  the time evolution of a single initial 
condition is enough to fills up the entire phase space. Finally, in case (iii), 
one can observe invariant tori,  chaotic seas generally surrounding 
Kolmogorov-Arnold-Moser  (KAM) coexisting. If a time dependent perturbation is 
introduced on the boundary \cite{td3}, the system exchanges energy with the 
moving particles upon collisions. Such type of systems have attracted a lot of 
attention lately because they can be used to study the phenomenon of unlimited 
energy growth also known as Fermi acceleration \cite{ref9}. However, how to 
identify in which type of system the phenomenon of Fermi acceleration will be 
observed? To answer this question,  Loskutov, Ryabov  and Akinshin (LRA) 
proposed a conjecture where  they state that a chaotic component in the phase 
space for the time-independent dynamics is a sufficient condition to observe 
Fermi acceleration once a time dependent perturbation on the boundary is 
introduced. This conjecture became known as LRA conjecture \cite{ref10,ref11} 
and over the years it has been verified in several systems such as the time 
dependent Lorentz gas \cite{ref12,ref12a}, oval \cite{ref13} and  stadium 
\cite{ref14,ref16} billiard among many other systems \cite{ref17,ref17a}. 
Nevertheless, later on results have shown that the existence of a chaotic 
component is a sufficient, but not a necessary condition for Fermi acceleration 
since the unlimited energy growth was also observed in a time dependent 
elliptical billiard. As it is known, the elliptical with static boundary is an 
integrable system whose integrability comes from the conservation of the angular 
momentum with respect to the two foci \cite{ref21}. However, when a time 
dependent perturbation is introduced into the system, the separatrix is replaced 
by a chaotic layer and trajectories that were confined inside the separatrix 
(librators) can now explore the region outside the separatrix (rotators) and 
vice-versa. This change of behavior, namely, librator orbits ``jumping'' to 
rotator and vice-versa turned out to be the mechanism which produces the 
unlimited energy growth  \cite{ref19,ref20,ref20a}. More recently, it became 
clear that such a phenomenon is not robust\cite{ref23} since a tiny 
amount of dissipation, either upon collision\cite{ref22} or during the flight 
\cite{ref23a}, is enough to suppress Fermi acceleration.

The motion of the time dependent boundaries can be related to a more physical 
situation. Indeed due to the thermal fluctuations, the position of each atom on 
the boundary is allowed to move locally. Such oscillation of the atoms, and 
hence of the boundary, can be extended to the context of billiard 
which allows us to connect the observables obtained from the velocity of the 
particle -- such as the kinetic energy -- to the thermodynamics, more 
precisely, the temperature and entropy. So far, such a connection has been made 
for the Lorentz gas to describe the motion of electrons between heavy ions as in 
a lattice of metal \cite{ref24}. Therefore, further investigation is needed to 
understand how the dynamics of the systems can be connected to more real 
situations.

In this paper we study some dynamical properties for an ensemble of 
noninteracting particles in a time dependent billiard. Our main goal is to 
understand and describe the dynamics of the mean squared velocity as a 
function of the time considering different values of the control parameters, 
namely, the dissipation upon collision, the parameter that controls the shape 
of 
the boundary and the amplitude of the time-dependent perturbation. 
of particles moving inside a closed domain. 
As an illustration, we consider a stochastic version of the time-dependent oval 
billiard. The introduction of a random perturbation on the border resembles the 
rough oscillations -- producing  a random exchange of energy upon collision -- 
at least in the microscopic domain. The results are obtained numerically are 
confirmed by using an analytical approach.

This paper is organized as follows. In Sec. \ref{sec2} we present the model. In 
Sec. \ref{sec4} we present our numerical findings. In Sec. \ref{sec3} we 
introduce an analytical approach to obtain the behavior of the average velocity 
as a function of the number of collisions with the moving boundary. Finally, in 
Sec. \ref{sec6}, we make a connection with the thermodynamics by using the 
expression obtained analytically for the particle's average velocity. 
Conclusions are drawn in Sec. \ref{sec7}.

\section{The model and the map}
\label{sec2}
\begin{figure}[t]
\centerline{\includegraphics[width=1\linewidth]{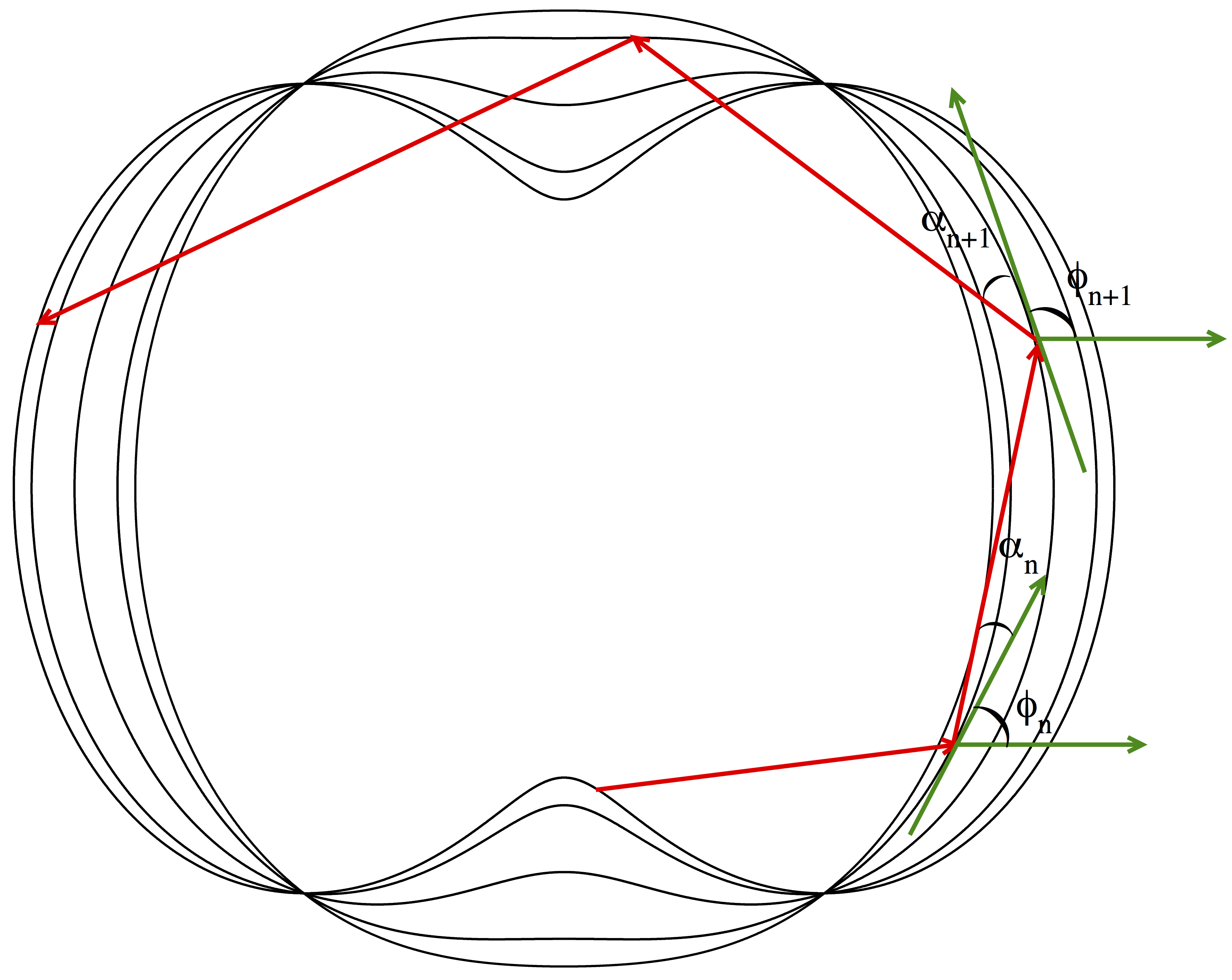}}
\caption{Illustration of 5 snapshots of a time-dependent oval billiard. The 
corresponding angles that describes the dynamics of the model is also shown for 
two collisions.}
\label{Fig1}
\end{figure}

In this section we present all the details needed to study the dynamics of an 
ensemble of noninteracting particles experiencing collisions with a moving 
boundary.  As it is so usual in the literature, we describe the dynamics of the 
model in terms of a four dimensional nonlinear mapping $\tilde{T}$ that gives 
the angular position of the particle $\theta_{n}$; the angle that the trajectory 
of the particle forms with the tangent line at the position of the collision 
$\alpha_{n}$; the absolute velocity of the particle $|\vec{V}_{n}|$ and the 
instant of the hit with the boundary $t_n$, i.e., 
$(\theta_{n+1},\alpha_{n+1},|\vec{V}_{n+1}|,t_{n+1})=\tilde{T}(\theta_{n}, 
\alpha_{n},|\vec{V}_{n}|,t_{n})$. The index $n$ denotes the $n^{th}$ collision 
with the moving boundary. Assuming that the  shape of the boundary in polar 
coordinates is $R_b(\theta,\epsilon,p,t)=1+\varepsilon f(t)\cos(p\theta)$ where 
the subindex $b$ denotes boundary, $f(t)$ is a function to be chosen, 
$\varepsilon$ is the perturbation of the circular billiard and $p$ is an integer 
number warranting a closed boundary, otherwise particles would escape. We 
consider $f(t)=1+\eta\cos(\omega t)$  where $\eta$ is the amplitude of 
the time dependent perturbation and $\omega$ is the angular frequency, which 
from now is fixed as $\omega=1$. Figure  \ref{Fig1} shows a typical illustration 
of a billiard and the angles used to describe the dynamics of the model.

For a given state $(\theta_n,\alpha_n,|\vec{V}_n|,t_n)$, the position 
of the particle as a function of the time is written as
\begin{eqnarray}
X(t)&=&X(\theta_n,t_n)+|\vec{V_n}|\cos(\alpha_n+\phi_n)(t-t_n),\label{eq1}
\\
Y(t)&=&Y(\theta_n,t_n)+|\vec{V_n}|\sin(\alpha_n+\phi_n)(t-t_n),
\label{eq2}
\end{eqnarray}
with $t\ge t_n$ and $X(\theta_n,t_n)=R_b(\theta_n,t_n)\cos(\theta_n)$ and 
$Y(\theta_n,t_n)=R_b(\theta_n,t_n)\sin(\theta_n)$. Furthermore, once the 
angular position $\theta$ is known, the angle between the tangent line and the 
horizontal at $X(\theta),Y(\theta)$ is 
$\phi=\arctan[Y^{\prime}(\theta,t)/X^{\prime}(\theta,t)]$ where 
$Y^{\prime}(\theta,t)=dY/d\theta$ and $X^{\prime}(\theta,t)=dX/d\theta$.  

Once the particle hits the moving boundary it is reflected and travels with a
constant speed until the particle experiences the  collision $(n+1)$. The 
distance traveled by the particle measured with respect to the origin of the 
coordinate system is given by $R(t)=\sqrt{X^2(t)+Y^2(t)}$. Therefore, the 
angular position at the next collision of the particle with the boundary, 
$\theta_{n+1}$, is obtained by solving the equation 
$R(\theta_{n+1},t_{n+1})=R_b(\theta_{n+1},t_{n+1})$. This means that the 
position of the particle is the same as the boundary, hence producing a 
collision. Moreover, we can also obtain the time at collision $n+1$ by 
evaluating the expression
\begin{eqnarray}
t_{n+1}=t_n+{{\sqrt{\Delta X^2+\Delta Y^2}} \over 
\vert\overrightarrow{V}_n\vert} +Z(n)~,
\label{eq6}
\end{eqnarray}
where $\Delta X=X_p(\theta_{n+1},t_{n+1})-X(\theta_{n},t_{n})$ and $\Delta 
Y=Y_p(\theta_{n+1},t_{n+1})-Y(\theta_{n},t_{n})$ and $Z(n)$ is a number taken 
at 
random from the interval $[0,2\pi]$ introduced after the instant of collision 
and it is the variable that introduces stochasticity into the model. To obtain 
the particle's velocity at time $t_{n+1}$, we should note that the referential 
frame of the boundary is moving. Furthermore, we assume the particle 
experiences 
a fractional loss of velocity/energy upon collision only on the normal 
component 
of the velocity. Therefore, at the instant of collision, the following 
conditions must be obeyed
\begin{eqnarray}
\vec{V^{\prime}}_{n+1}\cdot\vec{T}_{n+1}&=&\vec{V^{\prime}}_n\cdot 
\vec{T}_{n+1},\label{law1}\\
\vec{V^{\prime}}_{n+1}\cdot\vec{N}_{n+1}&=&-\gamma\vec{V^{\prime}}_n\cdot 
\vec{N}_{n+1},\label{law2}
\end{eqnarray}
where the unit tangent and normal vectors are
\begin{eqnarray}
\vec{T}_{n+1}&=&\cos(\phi_{n+1})\hat{i}+\sin(\phi_{n+1})\hat{j},\label{eq_t}\\
\vec{N}_{n+1}&=&-\sin(\phi_{n+1})\hat{i}+\cos(\phi_{n+1})\hat{j},\label{eq_nor}
\end{eqnarray}
The parameter $\gamma\in[0,1]$ is the restitution coefficient. Thus, if 
$\gamma=1$ we have the case of completely elastic collisions while $\gamma<1$ 
leads the particle to have partial loss of velocity/energy upon collisions. The 
term $\vec{V^{\prime}}$ corresponds to the velocity of the particle in the 
non-inertial reference frame. Using the equations above, one can find that the 
tangential and normal components of the velocity after collision $n+1$ are 
given by
\begin{eqnarray}
\vec{V}_{n+1}\cdot\vec{T}_{n+1}&=&\vec{V}_{n}\cdot\vec{T}_{n+1},\\
\vec{V}_{n+1}\cdot\vec{N}_{n+1}
&=&-\gamma\vec{V}_{n}\cdot\vec{N}_{n+1}\nonumber\\
&+&(1+\gamma)\vec{V}_{b}(t_{n+1}
)\cdot\vec{N}_{n+1},
\label{eq8b}
\end{eqnarray}
where $\vec{V}_{b}(t_{n+1})$ is the velocity of the boundary and it is written 
as 
\begin{eqnarray}
\vec{V}_{b}(t_{n+1})={dR(t)\over {dt}}\Big\vert_{t_{n+1}}
[\cos(\theta_{n+1})\widehat{i}+\sin(\theta_{n+1})\widehat{j}].
\label{eq10}
\end{eqnarray}

Finally, the  velocity of the particle after the collision $(n+1)$ is given by
\begin{eqnarray}
|\vec{V}_{n+1}|=\sqrt{(\vec{V}_{n+1}\cdot\vec{T}_{n+1}
)^2+(\vec{V}_{n+1}\cdot\vec{N}_{n+1})^2}
\label{eq012}
\end{eqnarray}
and $\alpha_{n+1}$ is written as
\begin{eqnarray}
\alpha_{n+1}=\arctan
\left[{\vec{V}_{n+1}\cdot\vec{N}_{n+1} \over
\vec{V}_{n+1}\cdot\vec{T}_{n+1}} \right]~.
\label{eq013}
\end{eqnarray}

With the equation above, we have all the ingredients needed to study the 
dynamics of the system.

\section{Numerical results for the stochastic dynamics}
\label{sec4}

\begin{figure}[t]
\centerline{\includegraphics[width=8.5cm,height=6cm]{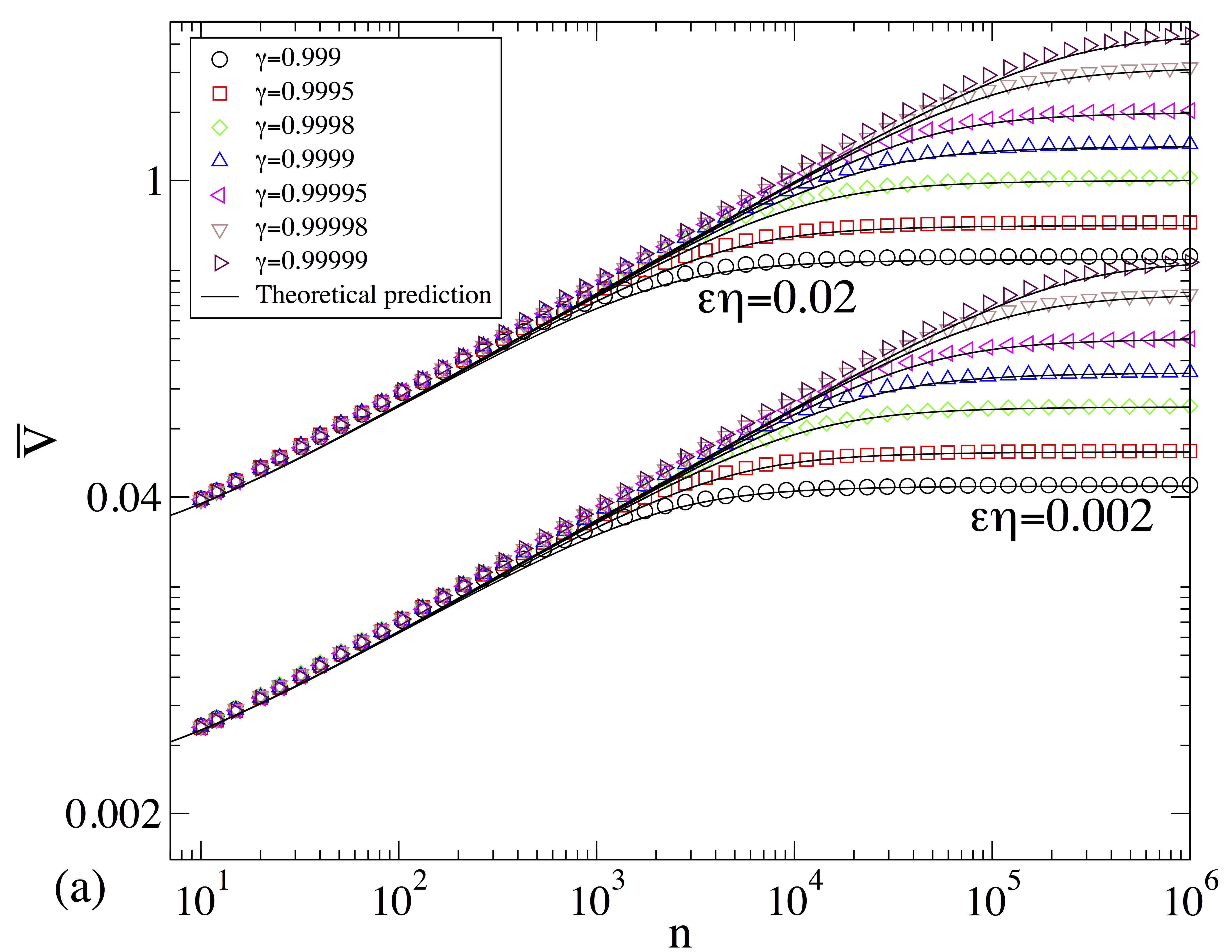}}
\includegraphics[width=8.5cm,height=6cm]{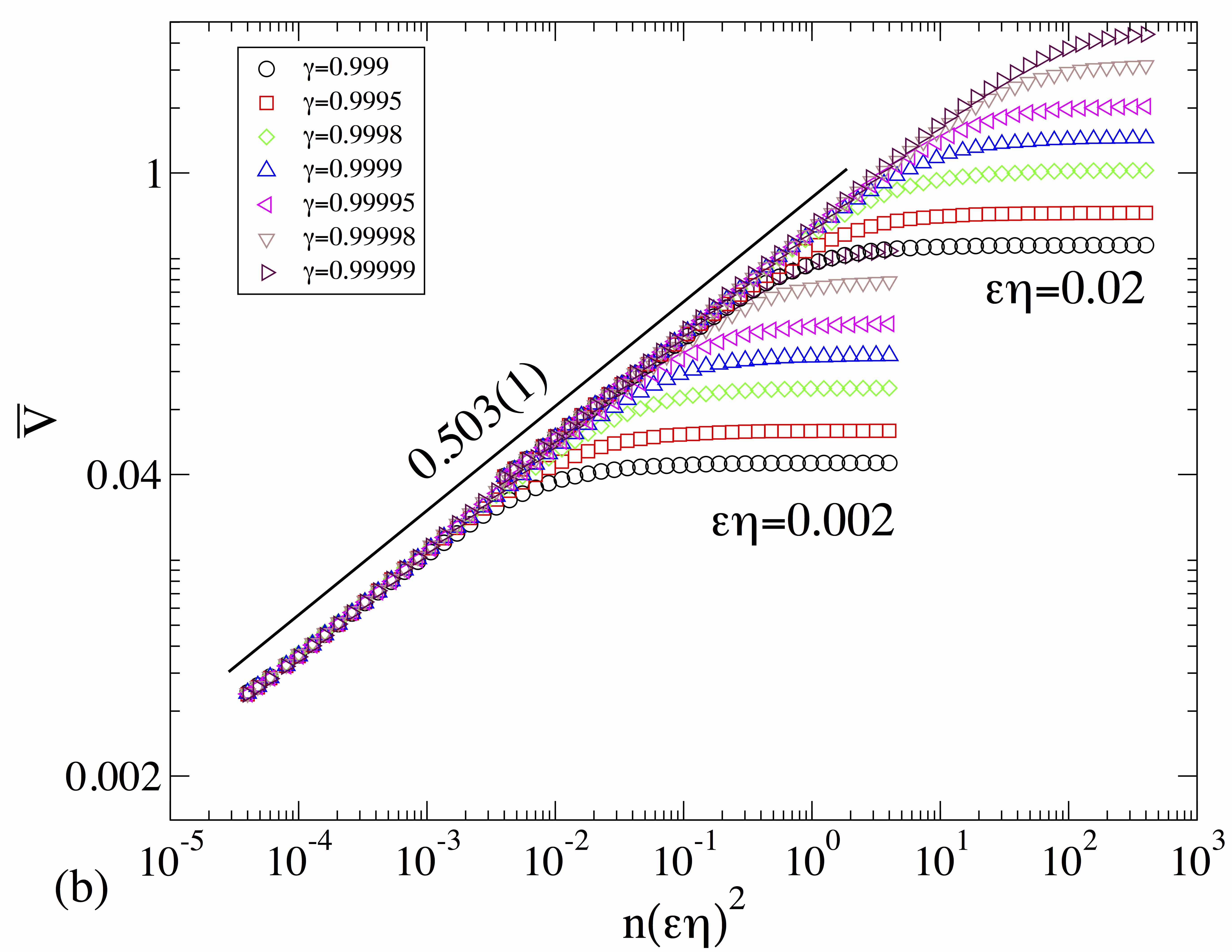}
\includegraphics[width=8.5cm,height=6cm]{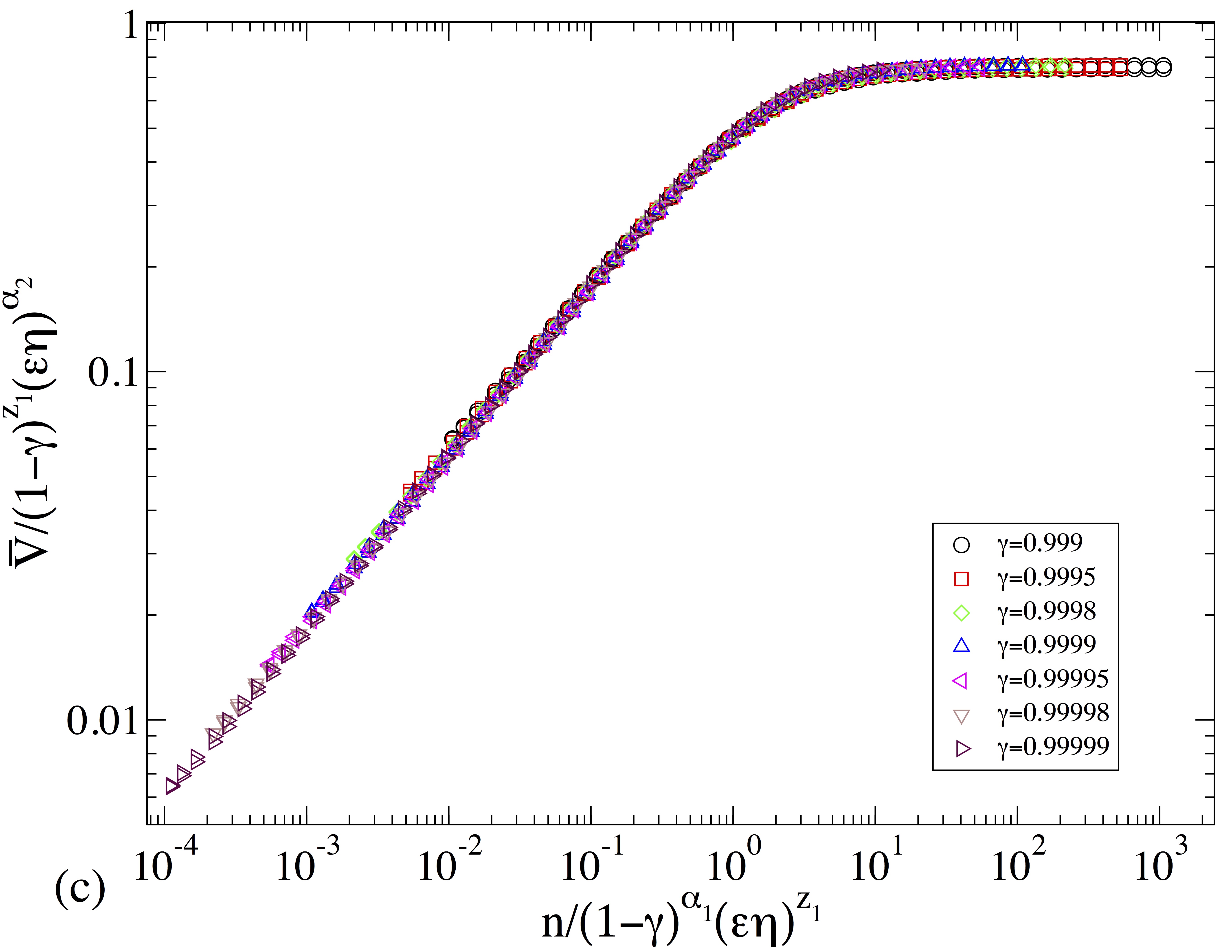}
\caption{(a) Behavior of the average velocity as a function of the number of 
collisions for different values of $\gamma$ and two different combination of 
$\eta\varepsilon$. (b) Their initial collapse after the transformation $n 
\rightarrow n(\varepsilon \eta)^2$. (c) Overlap of all the curves of the 
average velocity onto a single and universal plot.}
\label{Fig3a}
\end{figure}

\begin{figure}[t]
\centerline{\includegraphics[width=1.0\linewidth]{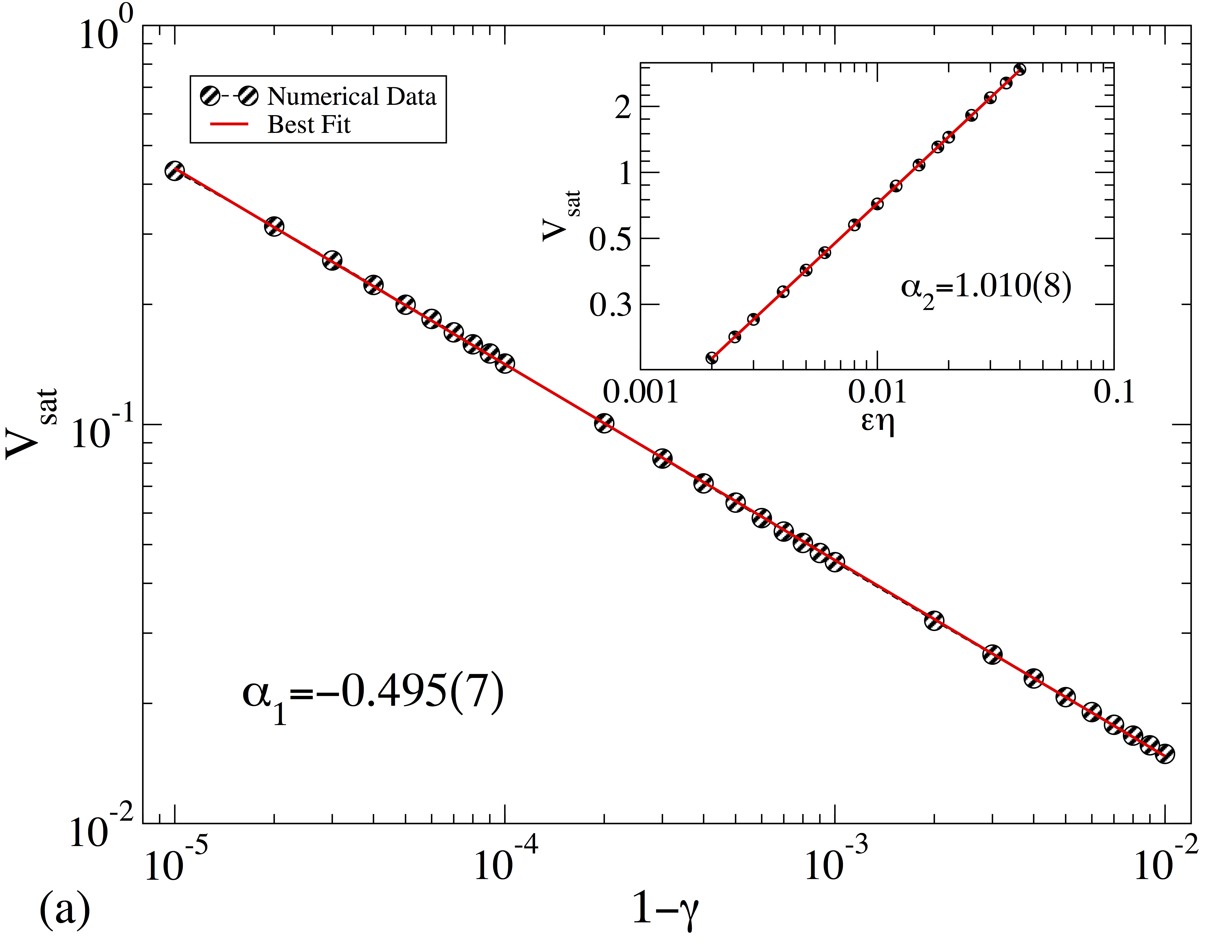}}
\includegraphics[width=1.0\linewidth]{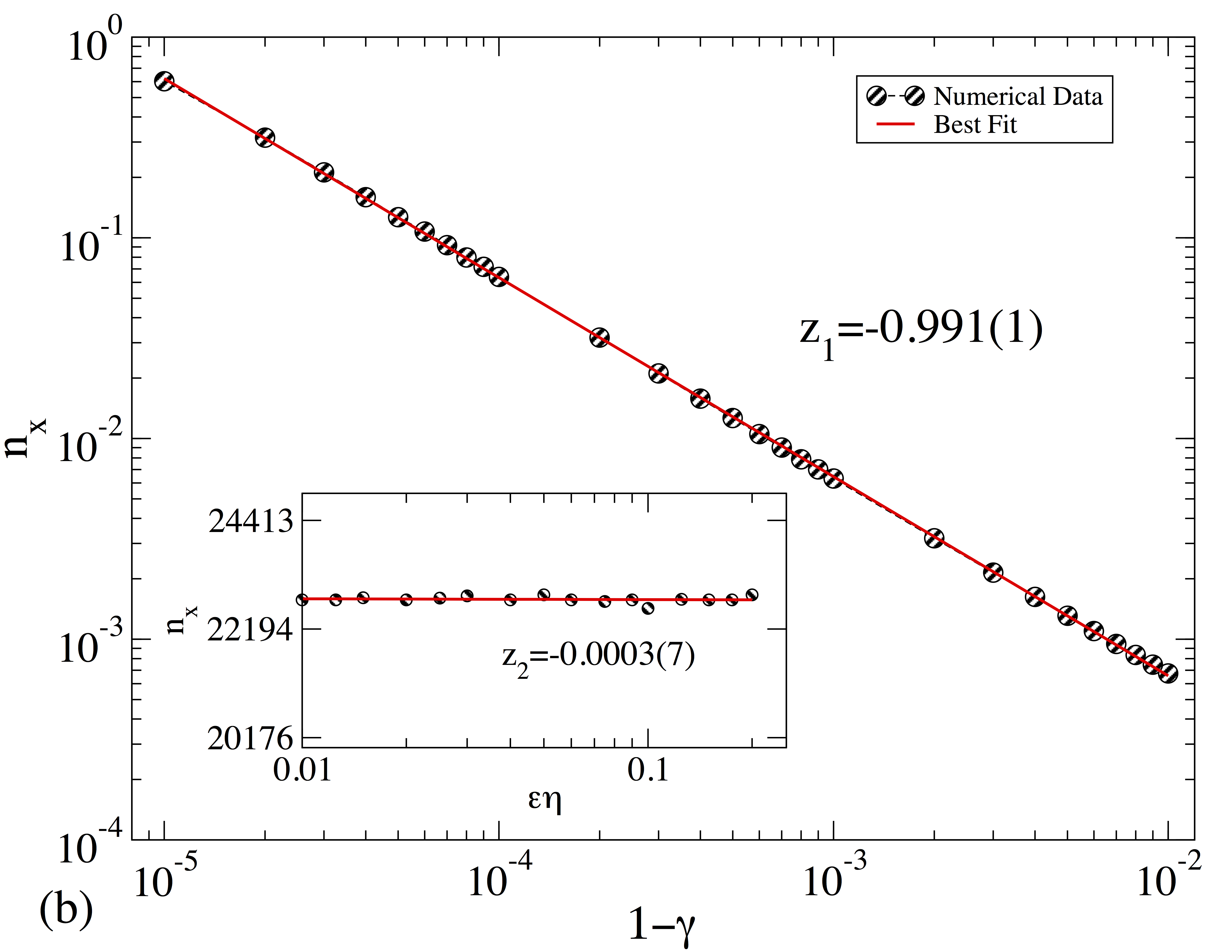}
\caption{Behavior of (a) $V_{sat}$ and (b) $n_x$ as a function of $(1-\gamma$). 
The inset shown the behavior of $V_{sat}$ and $n_x$ for different values of 
$\varepsilon\eta$. Observe that $n_x$ does not depend on $\varepsilon\eta$, 
hence $z_2=0$.}
\label{Fig3b}
\end{figure}

In this section we concentrate to characterize the behavior of the average 
velocity in terms of the number of collisions with the boundary and as a 
function of the control parameter $\gamma$, $\eta$ and $\varepsilon$. Here 
$\gamma$ is the dissipation parameter acting only along the normal component of 
the particle's velocity, $\varepsilon$ is the perturbation on the boundary and 
$\eta$ is the amplitude of the time perturbation. We consider a dissipative 
version of the oval-shaped billiard \cite{ref22} close to the transition 
from unlimited to limited energy growth. Such a transition happens when the 
control parameter $\gamma\rightarrow 1$. However, the transition is better 
characterized for $(1-\gamma)$. To obtain the average velocity, each initial 
condition has a fixed initial velocity, $V_0=10^{-3}$, $\eta\varepsilon \in 
[0.002,0.02]$ and randomly choose $\alpha_0 \in [0,\pi]$, $\theta_0 \in 
[0,2\pi]$, $t_0 \in [0,2\pi]$. In addition, after every time step, a random 
number $[Z(n)]$  is introduced in Eq. (\ref{eq6}), thus, introducing 
stochasticity into the model. Two different procedures were applied in order to 
obtain the average velocity. Firstly, we evaluate the average velocity over the 
orbit for a single initial condition and  then over an ensemble of initial 
conditions. Hence, the average velocity is written as
\begin{eqnarray}
\overline{V}={{1}\over{M}}\sum_{i=1}^M{{1}\over{n+1}}\sum_{j=0}^nV_{i,j}~,
\label{eq16}
\end{eqnarray}
where the index $i$ corresponds to a sample of an ensemble of initial 
conditions, $M$ denotes the number of different initial conditions. We have 
considered $M=2000$ in our simulations. The behavior of $\overline{V}~vs.~ n$ 
for different values of $\gamma$ is shown in Fig. \ref{Fig3a} (a). It is easy to 
see in Fig. \ref{Fig3a} two different kinds of behaviors. For short $n$, the 
average velocity grows according to a power law and suddenly it bends towards a 
regime of saturation for long enough values of $n$. The changeover from growth 
to the saturation is marked by a crossover iteration number $n_x$. Different 
values of $\eta\varepsilon$  generate different behaviors for the average 
velocity, however, applying the transformation $n \rightarrow 
n(\eta\varepsilon)^2$, all the curves start to grow together as shown in Fig. 
\ref{Fig3a} (b). For such a behavior, we propose the following scaling 
hypotheses:
\begin{enumerate}
\item{For small values of $n$, such as $n\ll n_{x}$, the growth regime can be 
described by
\begin{equation}
\overline{V}\propto[(\eta\varepsilon)^{2}n]^{\beta},
\label{a1}
\end{equation}
where $\beta$ is the acceleration exponent;
}
\item{For sufficiently large $n$, say $n \gg n_{x}$, where the regime of 
saturation is reached, we have
\begin{equation}
V_{sat}\propto(1-\gamma)^{\alpha_{1}}(\eta\varepsilon)^{\alpha_{2}},
\label{a2}
\end{equation} 
where $\alpha_{1}$ and $\alpha_{2}$ are the saturation exponents;
}
\item{Finally, the characteristic value of $n$ which provides the 
changeover from growth to the saturation is given by
\begin{equation}
n_{x}\propto(1-\gamma)^{z_{1}}(\eta\varepsilon)^{z_{2}},
\label{a3}
\end{equation}
where $z_{1}$ and $z_{2}$ are crossover exponents.
}
\end{enumerate}

Once the scaling hypotheses are known, we propose that the behavior of the 
average velocity can be described by a homogeneous function of the type
\begin{equation}
\overline{V}[(\eta\varepsilon)^{2}n,\eta\varepsilon,(1-\gamma)]=l\overline{V}[l^
{b}(\eta\varepsilon)^{2}n,l^{c}\eta\varepsilon,l^{d}(1-\gamma)],
\end{equation}
where $l$ is a scale factor, $b$, $c$ and $d$ are characteristic exponents that 
in principle must be related to the scaling exponents. Since $l$ is a scaling 
faction, one can choose the quantities $l^{b}(\eta\varepsilon)^{2}n=1$,  
$l^{c}\eta\varepsilon=1$ and $l^{d} (1-\gamma)=1$ and after some calculation 
\cite{ref26}, it is straightforward to show that the scaling exponents are 
related by the following expressions
\begin{eqnarray}
z_{1}={{\alpha_{1}}\over{\beta}},~~~~~~~~ 
z_{2}={{\alpha_{2}}\over{\beta}}-2.\label{sl_2}
\end{eqnarray}

Now, all we need  is to find the exponents numerically. A power law fitting in 
$\overline{V}$ when $n<<n_x$ gives us that $\beta=0.503(1)\simeq 1/2$. Such 
value was obtained from the range of $\gamma\in[0.999,0.99999]$. Considering 
$\eta\varepsilon$ fixed, after a power law fitting on the plot ${V}_{sat} 
~vs.~ (1-\gamma)$ we obtained $\alpha_1=-0.495(7)$ [see Fig. \ref{Fig3b} (a)]. 
The plot of ${n}_{x}~vs.~ (1-\gamma)$ gives us $z_1=-0.991(1)$. On the other 
hand, if $(1-\gamma)$ is fixed, the behavior of $V_{sat}~vs.~\eta\varepsilon$ 
gives a slope $\alpha_1=1.010(8)$ while a plot of $n_x~vs.~\eta\varepsilon$ 
furnishes $z_2=-0.0003(7)$. The crossover exponents $z_1$ and $z_2$ can be 
obtained also by using Eq. (\ref{sl_2}) and the previous values of $\alpha_1$, 
$\alpha_2$ and $\beta$. By doing so, we obtained $z_1=-0.98(1)$ and 
$z_2=0.00(1)$, which is in perfect agreement with the numerical data. A final 
check of the initial hypotheses can be seem in Fig. \ref{Fig3b} (c). As one can 
see in Fig. \ref{Fig3b} (c) all the curves overlap onto each other in a single 
and, hence, universal plot. Furthermore, the present result allows us to 
confirm that even in the stochastic dynamics, the introduction of dissipation is 
enough to suppress the unlimited energy growth. These results are confirmed by 
looking at Eq. (\ref{a2}) and Eq. (\ref{a3}) and the previous values for the 
critical exponents $\alpha_1$ and $z_1$. Observe that both $\alpha_1$ and $z_1$ 
are negative, which lead to $\overline{V}_{sat} \propto 1/(1-\gamma)^{\vert 
\alpha{_1} \vert}(\eta\varepsilon)^{\alpha_2} $ and $n_x \propto 
1/(1-\gamma)^{\vert z_1 \vert}(\eta\varepsilon)^{z_2} $. Note that when $ \gamma 
\rightarrow 1$, implies that $ \overline{V}_{sat} \rightarrow \infty $ and $ n_x 
\rightarrow \infty $, too, thus recovering the results for the conservative 
case leading then to Fermi acceleration. However, when $\gamma$ is slightly less 
than 1, it implies that the system has a characteristic saturation value $ 
\overline{V}_{sat}$ and a crossover iteration number $n_x$.

\section{Analytical approach }
\label{sec3}

Since the results obtained in the previous section were purely numerical, a 
more detailed analytical investigation is needed. To study analytically the 
average velocity of an ensemble of noninteracting particles, the corresponding 
probability distribution for the velocity in the two-dimensional phase space, 
i.e, $\alpha~vs.~\theta$, must be known. For the deterministic case, periodic 
islands are present in the phase space and obtain the probability distribution 
is a very difficult task when a time dependent perturbation is introduced. One 
possible way to overcome this limitation and to make the analytical work 
feasible is to assume that the probability distribution in the plane 
$\alpha~vs.~\theta$ is uniform. This is exactly what happens when the 
stochastic term is introduced in Eq. (\ref{eq6}). 

Taking this situation into account, we can average Eq. (\ref{eq012}) for the 
ranges $\theta\in[0,2\pi]$, $\alpha\in[0,\pi]$ and $t\in[0,2\pi]$. After some 
algebra we obtain
\begin{equation}
\overline{V^{2}}_{n+1}={{\overline{V^2}_n}\over{2}}+
{{\gamma^2\overline{V^2}_n}\over{{2}}}+{{(1+\gamma)^{2}\eta^{2}\varepsilon^{2}}
\over{8}}. 
\label{eq_v2}
\end{equation} 
This result will be used to discuss different regimes, namely, the stationary 
and the dynamical regime.

\subsection{Stationary state}

In the stationary regime the mean-squared velocity is obtained under the 
following condition 
$\overline{V^{2}}_{n+1}=\overline{V^{2}_{n}}=\overline{V^{2}}$, and that 
when isolating $\overline{V^{2}}$ leads to
\begin{eqnarray}
\overline{V^{2}}={{(1+\gamma)\eta^{2}\epsilon^{2}}\over{4(1-\gamma)}}.
\end{eqnarray}
If we define the root mean square velocity as 
$\overline{V}=\sqrt{\overline{V^{2}}}$, we have
\begin{equation}
\overline{V}={{\eta\epsilon}\over{2}}\sqrt{(1+\gamma)}(1-\gamma)^{-1/2}.
\label{eq_sat}
\end{equation}

From Eq. (\ref{eq_sat}) we see the exponent heading the term $(1-\gamma)$ is 
$-1/2$ which is in perfect agreement with the exponent $\alpha_1$ obtained 
from Fig. \ref{Fig3b} (a). We notice also that $\alpha_2=1$, in agreement also 
with the result from in the in-set of Fig. \ref{Fig3b} (a).

\subsection{Dynamical regime}

To investigate the behavior of the dynamical regime, the first thing we need 
to do is to transform a difference equation -- relating $\overline{V^{2}}_{n+1}$ 
with $\overline{V^{2}_{n}}$ -- from  Eq. (\ref{eq_v2}) into a differential 
equation whose solution is feasible. Then we assume that the average velocity, 
when averaged over a large ensemble, admits the following approximation
\begin{equation}
\overline{V^{2}}_{n+1}-\overline{V^{2}_{n}}={{\overline{V^{2}}_{n+1}-\overline{
V^{2}_{n}}}\over{(n+1)-n}}\cong {d{\overline{V^{2}}}\over{dn}},
\end{equation}
hence leading to
\begin{equation}
{d{\overline{V^{2}}}\over{dn}}={{\overline{V^{2}}}\over{2}}(\gamma^{2}-1)+{
{(1+\gamma)^{2}\eta^{2}\varepsilon^{2}}\over{8}}.
\end{equation}
Since this is a first order differential equation, it can be solved easily. 
Integrating the equation considering that at $n=0$ and  initial velocity $V_0$ 
we obtain
\begin{equation}
\overline{V^{2}}(n)=\overline{V^{2}_{0}}e^{{{(\gamma^2-1)}\over{2}}n}+ 
{{(1+\gamma)}\over{4(1-\gamma)}}\eta^{2}\varepsilon^{2}\bigg[1-e^{{{(\gamma^{2}
-1)}\over{2}}n}\bigg].
\label{v_din}
\end{equation}

When the initial velocity is sufficiently small, say $V_0\cong 0$, the dominant 
expression for $\overline{V^2}(n)$ is
\begin{equation}
\sqrt{\overline{V^2}(n)}={{(1+\gamma)^{1/2}}\over{2}}(1-\gamma)^{-1/2}
\eta\varepsilon 
\bigg[1-e^{{{(\gamma^{2}-1)}\over{2}}n}\bigg]^{1/2}.
\label{inicial}
\end{equation} 

On the other hand, if the initial velocity is large enough, the leading 
expression shows and exponential decay given by
\begin{equation}
\sqrt{\overline{V^2}(n)}\cong V_0e^{{{(\gamma^2-1)}\over{4}}n}\cong 
V_0e^{{{(\gamma-1)}\over{2}}n}.
\label{v_decay}
\end{equation}

Finally, the full dynamics of $\overline{V}(n)$ over an ensemble of initial 
condition is described by the following equation:
\begin{equation}
\sqrt{\overline{V^2}(n)}=\sqrt{\overline{V^{2}_{0}}e^{{{(\gamma^2-1)}\over{2}}n}
+{{(1+\gamma)}\over{4(1-\gamma)}}\eta^{2}\varepsilon^{2}\bigg[1-e^{{{(\gamma^{2}
-1)}\over{2}}n}\bigg]}.
\label{v_full}
\end{equation}

\subsection{Average over the orbit}

Since Eq. (\ref{v_full}) gives us the average velocity over an ensemble of 
initial conditions, we need to make an average over the orbit (or the 
number of collision) to compare the results with Fig. \ref{Fig3a}. Therefore we 
define a new velocity written as
\begin{equation}
V_{rms}=\sqrt{{{1}\over{n+1}}\sum_{j=0}^n\overline{V^2}(j)}.
\label{eq1_edl}
\end{equation}
The resulting velocity must be comparable with the one presented in Eq. 
(\ref{eq16}). After doing some algebra we end up with
\begin{widetext}
\begin{equation}
V_{rms}(n)=\sqrt{{{(1+\gamma)}\over{4(1-\gamma)}}\eta^{2}\varepsilon^{2}+{
1\over{(n+1)}}\Big[\overline{V^{2}_{0}}-{{(1+\gamma)}\over{4(1-\gamma)}}\eta^{2}
\varepsilon^{2}\Big]\Bigg[{{1-e^{{(\gamma^2-1)\over2}(n+1)}}\over{1-e^{
(\gamma^2-1)\over2}}} \Bigg]}.
\label{v_full2}
\end{equation}
\end{widetext}
From the above result one can see that
\begin{equation}
V_{rms}=\left\{\begin{array}{ll}
V_0, ~~~~~~~~~~~~~~~~~~~~~~~~~~~~~ if~~~n=0\\
{{\eta\varepsilon}\over{2}}\sqrt{(1+\gamma)}(1-\gamma)^{-1/2}, ~~~ if ~~~ 
n\rightarrow\infty\\
\end{array}
\right.~.
\label{eq001}
\end{equation}

The continuous line in Figure \ref{Fig3a} (a) shows the behavior of the 
theoretical value of the square velocity [Eq. \ref{v_full2}] as a function of 
the number of collisions for different combinations of control parameters 
$\gamma$ and different values of $\eta\varepsilon$. As one can see, it reproduces 
remarkably well the results obtained numerically.

\subsection{Scaling exponents}

To obtain the scaling exponents one must consider three situations: (i) at the 
saturation, when $n\gg n_x$, $\overline{V}$ is described by Eq. (\ref{eq001}) 
and after comparing it with Eq. (\ref{a1}) one can see that $\alpha_1=-1/2$ and 
$\alpha_2=1$. On the other hand, for small values of $n$ such as $n\ll n_x$, 
the mean-square velocity behaves according to Eq. (\ref{v_full2}). Assuming 
$V_0<<1$, and expanding the exponential in a Taylor series 
\begin{equation}
e^{{{(\gamma^{2}-1)(n+1)}\over{2}}}=1+{{(\gamma^{2}-1)(n+1)}\over{4}}+\ldots,
\end{equation} 
and considering $(n+1)\cong n$ and taking only the lower order in the 
expansion, we can rewrite $\overline{V}$ as
\begin{equation}
\overline{V}(n)={{(1+\gamma)}\over{4}}(\eta\varepsilon)n^{1/2},
\label{eq_n}
\end{equation}
Therefore, comparing Eq. (\ref{eq_n}) with Eq. (\ref{a1}) one can show that the 
acceleration exponent is $\beta=1/2$. Finally, the crossover iteration number 
$n_x$ that marks that change of behavior from growth to saturation can be 
estimated when Eqs. (\ref{eq_n}) intersect  Eq. (\ref{eq001}). After some 
calculation we obtained
\begin{equation}
n_{x}={{4}\over{(1+\gamma)}}(1-\gamma)^{-1}.
\end{equation}
After a comparison with the scaling hypotheses (\ref{a3}), we can show that the 
saturation exponents are $z_1=-1$ and $z_2=0$. The exponents $z_1$ and $z_2$ can 
be obtained also by using Eq. (\ref{sl_2}) and the previous values of $\beta$, 
$\alpha_1$ and $\alpha_2$.

\section{Connection with Thermodynamics}
\label{sec6}
\begin{figure}[t]
\centerline{\includegraphics[width=1\linewidth]{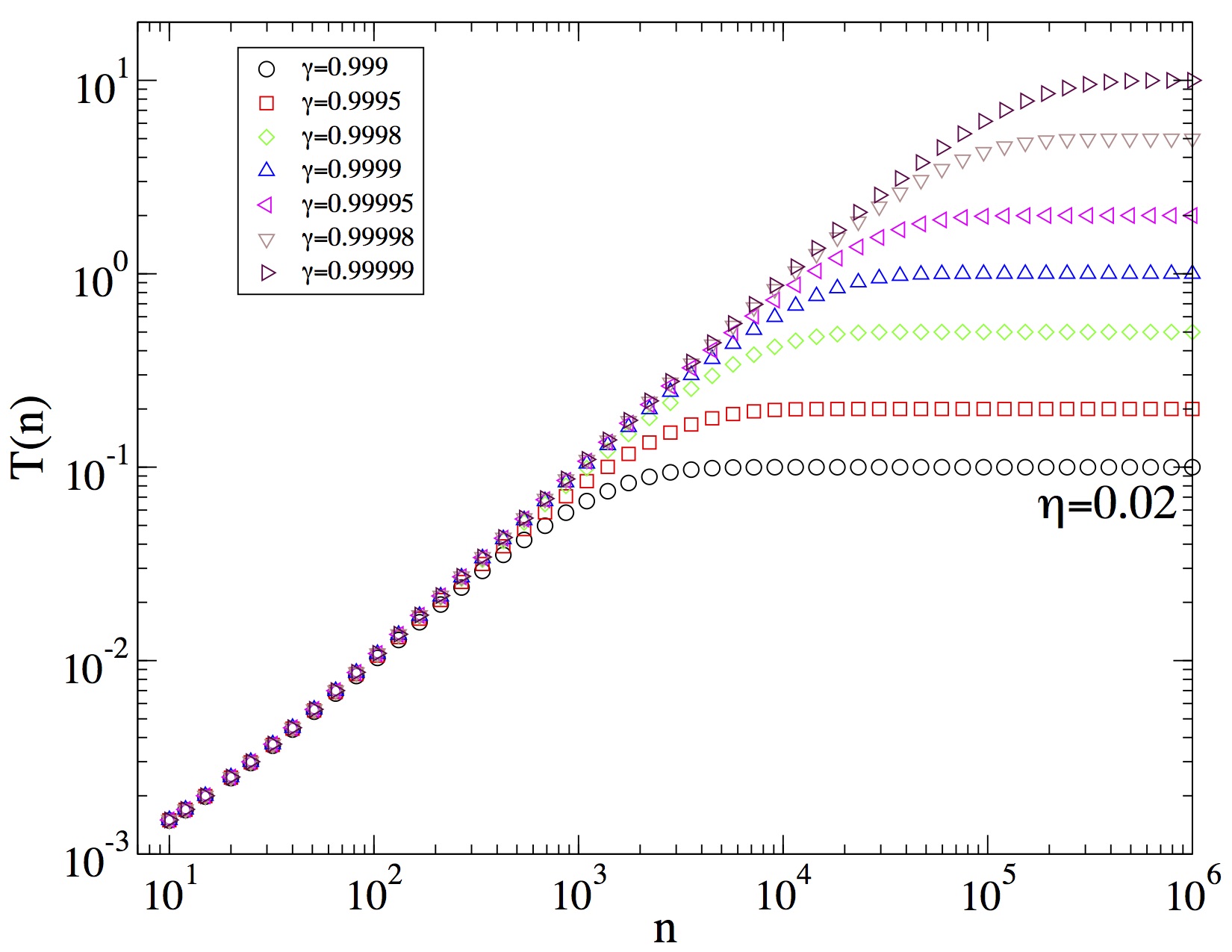}}
\caption{Behavior of the temperature $T$ as a function of $n$.}
\label{Fig4}
\end{figure}

In this section, let us discuss a possible connection with the thermodynamic. 
To do so, we make use of the energy equipartition theorem \cite{ref26,ref25} 
which can be given in a compact form as
\begin{equation}
{{m}\over{2}}\overline{V^{2}}=K_{B}T=U,
\end{equation}
where $K_B$ is the Boltzmann constant and $U$ is the energy of the gas of 
particles. Once the squared velocity is known, the temperature can be written as
\begin{equation}
T(n)=T_0e^{[{{(\gamma^2-1)n}\over{2}}]}+{{m}\over{2K_B}}
{{(1+\gamma)}\over{4(1-\gamma)}}\eta^{2}\varepsilon^{2}[1-e^{[{{(\gamma^{2}
-1)n}\over{2}}]}].
\end{equation}
If the ensemble of particles in the billiard is started with low temperature 
$T_0\ll{{m}\over{2K_B}}{{(1+\gamma)}\over{4(1-\gamma)}}\eta^{2}\varepsilon^{2}$, 
then the temperature will behave as in Fig. \ref{Fig4}, exhibiting similar 
scaling features as seen in Fig. \ref{Fig3a}.

An expression for the entropy, as foreseen from statistical mechanics standpoint 
in differential form, then
\begin{equation}
dS={{1}\over{T}}dU.
\end{equation}
Integrating this equation we end up with
\begin{equation}
S=\tilde{S} + K_{b}\ln(U),
\end{equation}
where $\tilde{S}=S_{0} -K_{b}\ln(U_{0})$, $S_0$ is a constant and $U_0$ is the 
initial energy. We see that the entropy confirms a classical postulate from the 
thermodynamics \cite{ref27} that it is a monotonically growing function of the 
energy.

\section{Conclusion}
\label{sec7}

As the concluding remark, we have studied some dynamical properties of a 
dissipative time dependent oval billiard. We introduced a random perturbation on 
the boundary and investigate the behavior of the average velocity of the 
particles as a function of the number of collisions for different combinations 
of the control parameters. We observed that average velocity grows for small 
number of collision and then, after a crossover, it reaches a regime of 
saturation for large $n$. Thus we do not observe the unlimited energy growth 
(Fermi acceleration). We have shown there is a relationship between the 
critical exponents $\beta$, $\alpha_1$, $\alpha_2$, $z_1$ and $z_2$. The 
exponents were used to show that the average squared velocity is scaling 
invariant with a perfect collapse of all curves onto a single universal plot. 
Additionally, we have given an analytical argument using the equilibrium 
condition at the steady state regime for the scaling exponents obtained 
numerically. Our procedure led to obtain the critical exponents $\beta=0.5$, 
$\alpha_1=-0.5$, $\alpha_2=1$, $z_1=-1$ and $z_2=0$ which are in perfect 
agreement with the exponents obtained numerically. The empirical expression for 
the average velocity allowed us to make a connection with the thermodynamic. Our 
results show that the average squared velocity -- and hence the temperature -- 
grows as a power law for $V_0\sim 0$, but and $n \rightarrow \infty$, it tends 
to a constant plateau. Furthermore, the weaker is the dissipation, the higher if 
the temperature in the steady state.  Finally, we confirmed that the entropy is 
monotonically growing function of the energy.

\acknowledgments
MVCG acknowledges CNPq for financial support. DFMO thanks to James S. McDonnell 
Foundation. EDL thanks FAPESP (2012/23688-5) and CNPq (303707/2015-1) for 
financial support.

\end{document}